\documentclass[%
  reprint,
  superscriptaddress,
  amsmath,amssymb,
  aps,
superscriptaddress,
  prl,
longbibliography,
]{revtex4-1}

\usepackage{xcolor}
\usepackage{physics}
\usepackage{graphicx}
\usepackage{dcolumn}
\usepackage{bm}
\usepackage{siunitx}
\usepackage[normalem]{ulem}

\newcommand{\supp}{(Supplemental Material)}
\newcommand{\ECS}{E_\mathrm{C}^\mathrm{S}}
\newcommand{\ECD}{E_\mathrm{C}^\mathrm{QD}}

\newcommand{\new}[1]{{\color{black}#1}}
\newcommand{\secondreview}[1]{{\color{black}#1}}

\renewcommand{\sout}[1]{}

\begin{document}

\title{Revealing charge-tunneling processes between a quantum dot and a superconducting island through gate sensing}

\author{Jasper van Veen}
\affiliation{QuTech and Kavli Institute of Nanoscience, Delft University of Technology, 2600 GA Delft, The Netherlands}

\author{Damaz de Jong}
\affiliation{QuTech and Kavli Institute of Nanoscience, Delft University of Technology, 2600 GA Delft, The Netherlands}

\author{Lin Han}
\affiliation{QuTech and Kavli Institute of Nanoscience, Delft University of Technology, 2600 GA Delft, The Netherlands}

\author{Christian Prosko}
\affiliation{QuTech and Kavli Institute of Nanoscience, Delft University of Technology, 2600 GA Delft, The Netherlands}

\author{Peter Krogstrup}
\affiliation{Center for Quantum Devices, Niels Bohr Institute, University of Copenhagen \& Microsoft Quantum Materials Lab Copenhagen, Denmark}

\author{John D. Watson}
\affiliation{Microsoft Quantum Lab Delft, Delft University of Technology, 2600 GA Delft, The Netherlands}

\author{Leo P. Kouwenhoven}
\affiliation{QuTech and Kavli Institute of Nanoscience, Delft University of Technology, 2600 GA Delft, The Netherlands}
\affiliation{Microsoft Quantum Lab Delft, Delft University of Technology, 2600 GA Delft, The Netherlands}

\author{Wolfgang Pfaff}
\email{wolfgang.pfaff@microsoft.com}
\affiliation{Microsoft Quantum Lab Delft, Delft University of Technology, 2600 GA Delft, The Netherlands}

\date{\today}

\begin{abstract}
We report direct detection of charge-tunneling between a quantum dot and a superconducting island through radio-frequency gate sensing.
We are able to resolve spin-dependent quasiparticle tunneling as well as two-particle tunneling involving Cooper pairs.
\new{The quantum dot can act as an RF-only sensor to characterize the superconductor addition spectrum, enabling us to access subgap states without transport.}
Our results provide guidance for future dispersive parity measurements of Majorana modes, which can be realized by detecting the parity-dependent tunneling between dots and islands.
\end{abstract}

\maketitle

Quantum dots coupled to superconductors can give rise to novel physical phenomena such as $\pi$ and $\phi_0$-junctions \cite{de2010hybrid, van2006supercurrent, szombati2016josephson}, Cooper pair splitting \cite{hofstetter2009cooper, herrmann2010carbon, deacon2015cooper}, and Yu-Shiba-Rusinov (YSR) states \cite{ralph1995spectroscopic, lee2014spin}. 
These phenomena arise because the single-electron states of the dot hybridize with the more complicated many-particle states of the superconductor.
Recently, such hybrid systems have gained interest in the context of Majorana zero modes (MZMs) where the quantum dot (QD) can, for example, be used as a spectrometer \cite{deng2016majorana}. 
Moreover, projective parity measurements can be achieved by coupling a QD to a pair of MZMs, which are located on a superconducting island (SC) \cite{higginbotham2015parity, albrecht2016exponential}, enabling topologically protected quantum computation.
These projective measurements rely on the parity-dependent hybridization between a single dot level and the MZMs \cite{plugge2017majorana, karzig2017scalable}.
Therefore, unambiguous detection of coherent tunneling between a QD and a superconducting island is needed to implement this readout. 

Dispersive gate sensing \new{(DGS)} provides direct access to charge hybridization between weakly coupled dots or islands.
More precisely, \sout{coherent}tunneling within these structures can impart a frequency shift on a resonant circuit that can be observed on short time scales with high accuracy.
In this way, experiments have revealed coherent charge hybridization between superconductors \cite{wallraff2004strong, duty2005observation, esmail2017cooper} and semiconductor double quantum dots \cite{petersson2012circuit, frey2012dipole, colless2013dispersive, landig2019microwave}.
Moreover, capacitive RF sensing has been used to study charging of QDs connected to normal and superconducting reservoirs \cite{gonzalez2015probing, bruhat2016cavity, bruhat2018circuit}.
\new{As such, DGS presents an excellent opportunity for studying charge-tunneling in hybrid structures containing QDs.}

In this paper, we report \new{direct} detection \sout{and identification}\new{of different types} of charge-tunneling processes between a QD and a SC through \sout{RF \new{gate} sensing}\new{DGS} \sout{via an LC resonator connected to the gate of}on the QD. 
From observations of the resonator response, supported by numerical simulations\sout{ of the system}, we find that the nature of the tunneling depends crucially on the ordering of the relevant energy scales of the SC. 
When the smallest scale is the energy of the lowest single-particle state, the QD and SC can exchange quasiparticles, giving rise to \new{the} characteristic `even-odd' effect.
Conversely, when the charging energy of the SC is lowest, we detect signatures of Cooper pairs tunneling out of the SC.
Depending on the tunneling amplitude, this results in either 1$e$-charging of the QD, with the other electron leaving into a reservoir, or 2$e$-charging of the QD\sout{via coherent Cooper pair tunneling}.
\sout{We can re-enable the}Tunneling to the single-particle states can, however, be re-enabled by operating the device in a floating regime where the total number of charges in the two systems is conserved.
\new{These results show that DGS allows us to effectively perform RF-only tunneling spectroscopy on the SC.
To this end, we use the QD and \secondreview{capacitively-coupled} resonator as a probe to characterize a subgap state in the SC without need for transport via leads.}
\secondreview{Our method is complementary to recent experiments that employed the dispersive response of inductively-coupled resonators to probe the Andreev bound state occupation in galvanically-isolated nanowire Josephson junctions \cite{hays2018direct, tosi2019spin}.}

\begin{figure}[ht!]
	\includegraphics[width=0.5\textwidth]{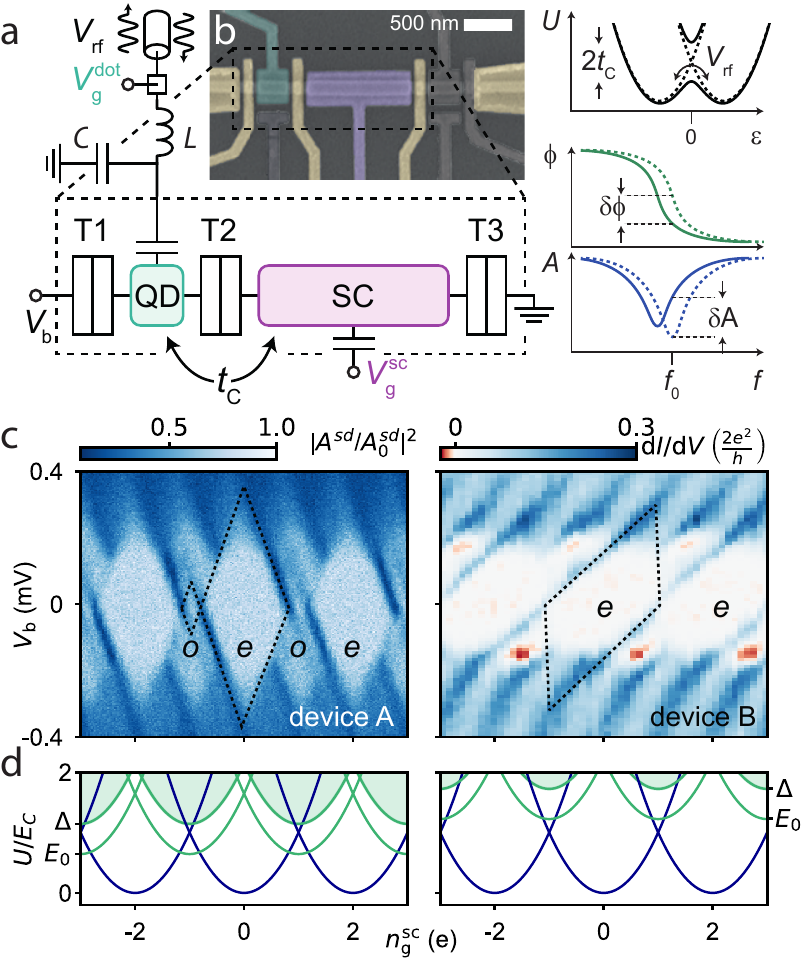}
	\caption{
	    \textbf{Experimental setup and sample characterization.}
	    \textbf{a)} \new{A gate sensor detects the tunneling of charges on/off the QD. 
	    Charge hybridization between QD and SC results in an anti-crossing in the energy spectrum at zero detuning $\epsilon=0$, and shifts the $LC$ resonator, causing in a change in amplitude, $A$, and phase, $\phi$, response of a probe signal at the bare resonator frequency, $f_0$.} 
	    \textbf{b)} False-colored electron micrograph of a nominally equivalent hybrid double dot.
        \new{An additional dot, shown on the right of the device, is unused in this experiment.}
        \textbf{c)} Coulomb blockade measurement of the SC. Left: device A, measured using RF reflectometry off the source (circuit not shown). The even-odd pattern indicates $E_0<\ECS$. Right: device B, measured using standard lock-in techniques. The doubling of the period at low bias $V_\text{b}$ illustrates that $E_0>\ECS$.
        \textbf{d)} Energy dispersion of the superconducting island for device A (left) and device B (right). The even (odd) energy levels are shown in blue (green).
        The odd parity sector consists of a discrete subgap state at $E_0$ and a continuum of states above $\Delta$ (shaded).
    } 
	\label{Fig1}
\end{figure}

A schematic of our experiment is shown in Fig.~1a,b. 
\secondreview{The QD-SC hybrid double dot} is formed in an InAs nanowire with an epitaxially grown Al-shell \secondreview{on two of its facets}.  
\secondreview{The superconducting island consists of a proximitized wire segment, which is defined by removing the Al outside a \SI{1.2}{\micro\meter} window using wet etching.}
\secondreview{The low-carrier density in the wire allows for gate-tunable subgap states in the SC \cite{antipov2018effects}.}
Tunneling barriers are implemented with gates, insulated from the wire by 10\,nm AlO$_x$. 
They are used to define the QD and to control the various tunneling rates. 
Large-lever arm top gates (`plungers') on both QD and SC can be used to tune the chemical potentials.
\secondreview{We have measured two nominally-identical devices, labelled A and B.}
\secondreview{For both devices, we connected the QD} plunger to an off-chip, superconducting resonator \new{with a resonance frequency of 449.5\,(443.2)\,MHz for device A (B)} \supp\,\cite{hornibrook2014frequency}.
We use its response near the resonance frequency to probe charge tunneling on and off the dot.
\new{All measurements are done at temperatures of $T \approx$ 20\,mK and at zero magnetic field unless otherwise indicated.}
\sout{We have fabricated two of these devices, and measured them separately at temperatures of $T \approx$ 20\,mK \sout{in a dilution refrigerator}\new{at zero magnetic field}.}

The relevant energy scales in our devices can be obtained from Coulomb blockade measurements:
Figure~1c  shows Coulomb diamonds of the superconducting island alone, measured through conductance. 
The diamonds of device A display a clear even-odd pattern, indicating that the energy of the lowest odd-parity state, $E_0$, is smaller than the charging energy of the superconducting island, \new{$\ECS=e^2/2C_\Sigma^\text{S}$, where $C_\Sigma^\text{S}$ is the total capacitance of the SC} (Fig.~1d) \cite{averin1992single,van2016conductance}.
For this device, we estimate $E_0=\SI{72}{\micro\electronvolt}$ and $\ECS=\SI{112}{\micro\electronvolt}$ from the extent of the diamonds.
Conversely, the charging of the superconducting island of device B is 2$e$-periodic, indicating that $E_0 > \ECS$ \cite{hekking1993coulomb,van2016conductance};
here, we estimate $E_0 \approx \SI{90}{\micro\electronvolt}$ and $\ECS \approx \SI{70}{\micro\electronvolt}$.
While in an ideal BCS superconductor $E_0$ is equal to the superconducting gap $\Delta$, current measurements on device A \supp\ and the negative differential conductance observed in device B indicate the presence of subgap states \cite{higginbotham2015parity}.
\secondreview{We attribute the difference in charging energies and $E_0$ between the devices to a combination of a slightly smaller wire diameter for device A and typical sample-to-sample variations arising from fabrication.}
In both devices, the charging energy of the dot, $\ECD\approx200-\SI{300}{\micro\electronvolt}$, is the largest energy scale in the system, and the typical QD level spacing exceeds the thermal energy \supp.

\begin{figure}[t!]
	\includegraphics[width=0.5\textwidth]{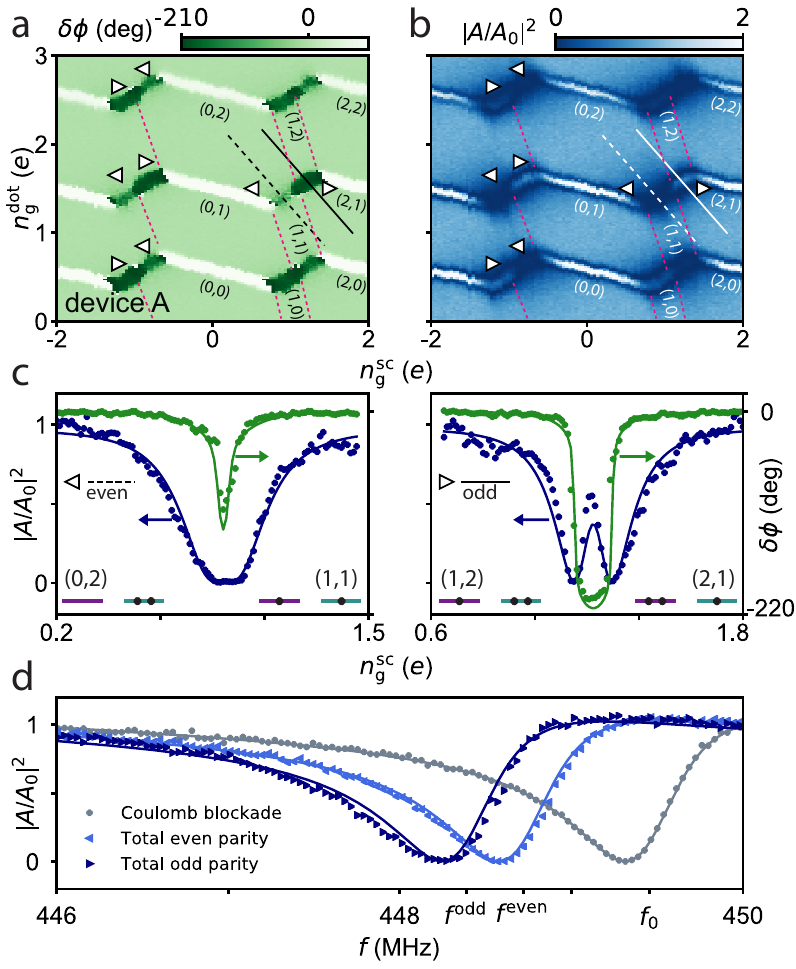}
	\caption{
        \textbf{Spin-dependent tunneling between a QD and a SC.}
        \textbf{a)} and \textbf{b)} Charge stability diagram of device A measured in phase \textbf{a)} and amplitude \textbf{b)}. 
        The charge states are labeled $\left(n^{\text{SC}},n^\text{QD}\right)$ with respect to the state $(N,M)$ with $N$ and $M$ even. 
        Dashed pink lines: expected locations of the lead-island transitions, \new{in accordance with data obtained from a resonator connected to the SC gate, and simulations of the charge ground state of the system (Supplemental Material)}.
        \textbf{c)} Linecuts of the phase (green \secondreview{dots}) and amplitude (blue \secondreview{dots}) along two interdot transitions. 
        Dashed line and left-pointing triangle marker, left panel: transition between (0,2) and (1,1), representative of  transitions between states of total even parity.
        Continuous line and right-pointing triangle marker, right panel: transition between (1,2) and (2,1), representative of transitions between states of total odd parity.
        \secondreview{The lines are fits to a circuit QED model with $t^\text{odd}_\text{C}=t^\text{even}_\text{C}/\sqrt{2}$ \supp. 
        We find a good fit with $t^\text{even}_\text{C}=\SI{20}{\giga\hertz}$ and $g_0=\SI{100}{\mega\hertz}$.}
        \new{\textbf{d)} Full-frequency response of the resonator (symbols) together with fits (lines) obtained from a pair of interdot transitions that show the parity effect (outside the gate space shown in \textbf{a,b}).
        } 
    }
	\label{Fig2}
\end{figure}

\begin{figure*}
	\includegraphics{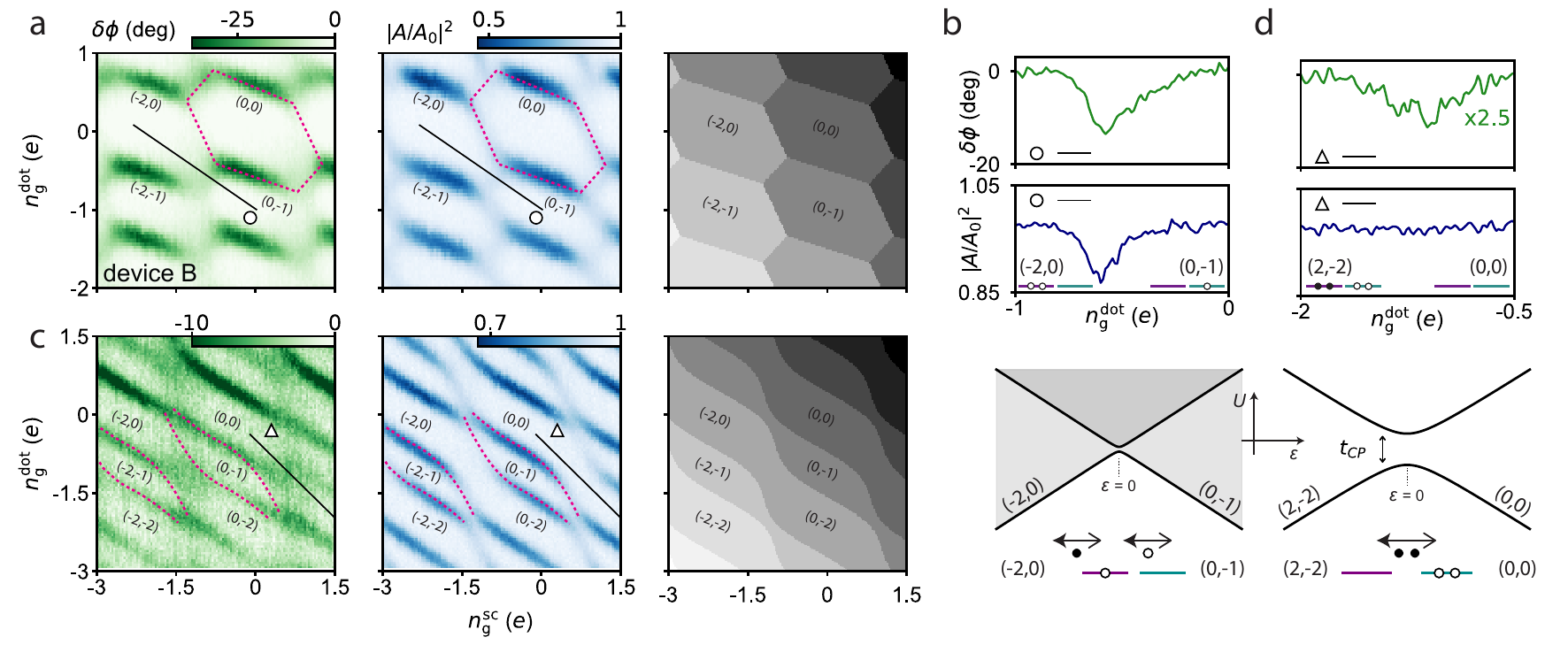}
	\caption{
        \textbf{Cooper pair tunneling in a hybrid double dot.}
        \textbf{a)} Charge stability diagram measured in phase (left) and amplitude (middle) along with a simulation of the charge ground state (right) in the weakly coupled regime. 
        The charge states are labeled $\left(n^{\text{SC}},n^\text{QD}\right)$ with respect to the state $(N,M)$ with $N$ even.
        Dashed pink lines: locations of the transitions from the (0,0) state as a guide to the eye. The gray scale in the simulation indicates the sum of the charge in the combined system.
        \textbf{b)} Linecuts of the phase (green) and amplitude (blue) along the (-2,0) to (0,-1) interdot transition. 
        This transition involves a reservoir with a continuous spectrum, indicated by the shaded region above the lowest available energy state.
        The schematic shows how these states couple via crossed Andreev reflection.
        \textbf{c)} Same as in \textbf{a)} for the strongly coupled regime. 
        Dashed pink lines: locations of the lead transitions from the (-2,-1) and (0,-1) states as a guide to the eye.
        \textbf{d)} Linecuts of the phase (green) and amplitude (blue) along the (2,-2) to (0,0) interdot transition. 
        These states couple via \sout{coherent}Cooper pair tunneling. 
        \sout{All data is measured in device B.}
    }
	\label{Fig3}
\end{figure*}

In the following, we investigate the change in resonator response when charges are able to tunnel between the QD and SC at zero bias, beginning with device A.
To this end, we form a hybrid double dot by tuning the gates T1 and T2 close to pinch-off, and T3 into pinch-off.
Figures~2a,b show the resonator response as a function of the two plunger gates in the weakly coupled regime.
Both amplitude and phase response display the charge stability diagram (CSD) of the hybrid double dot, which shows a clear 1$e$ pattern along the QD gate, and an even-odd pattern along the SC gate; 
this is again a manifestation of $E_0 < \ECS$, and the CSD shape can be readily reproduced by computing the charge ground states of the system \supp.
 
We focus on the interdot transitions, highlighted in Figs.~2a-c, where we observe a strong amplitude and phase response on all charge degeneracy points.
Interestingly, we see a strong difference in the resonator response across interdot transitions with a different parity of the total particle number, indicating a difference between the coupling between the involved states \cite{esterli2018small}.
\sout{Two scenarios can lead to such a different coupling:
One, an asymmetric electron- and hole coupling to the quasiparticle state in the SC \cite{hansen2018probing}; and second, a difference in the available spin states for the different transitions \cite{cottet2011mesoscopic}.}

\new{This dependence on total parity can be understood as a spin-dependent tunnel coupling \cite{cottet2011mesoscopic}.} \sout{The latter situation can qualitatively describe the asymmetry in our data.}To see this, we label the states according to their pairing; for the SC states as even/odd, and for the QD states as singlet/doublet: $\ket{e/o, S/D}$.
We can differentiate couplings between two sets of states; $\ket{e, D}$ to $\ket{o, S}$ and $\ket{e, S}$ to $\ket{o, D}$. 
\new{The coupling is different for these two sets because of spin degeneracy:
From $\ket{e, S}$ either of the two electrons forming the singlet can tunnel with equal probability, giving rise to a two-fold degeneracy in the coupling between $\ket{e, S}$ and $\ket{o, D}$, i.e., between the states with total even parity.
On the other hand, from $\ket{e, D}$ only one of the two electrons can tunnel from QD to SC due to Pauli exclusion; in this case no degeneracy is present. 
As a result, the transition rate between states of total even parity is expected to be $\sqrt{2}$ times stronger than between states of total odd parity, corresponding to a $\sqrt{2}$ times smaller frequency shift.

\secondreview{
We model the resonator response across the interdot transition by using a circuit QED model derived from input-output theory \supp\, \cite{petersson2012circuit}.
We find a good agreement between our data and the model using $t^\text{odd}_\text{C}=t^\text{even}_\text{C}/\sqrt{2}$.
However, the weak parameter dependence of the model makes it unreliable to confirm the spin-dependence of the tunnel coupling.}
To obtain quantitative agreement, we measure the frequency-dependent resonator response at a pair of interdot transitions that show this parity effect (Fig.~2d).
From fits to the resonator responses we extract frequency shifts with a ratio of $(f_0-f^\text{odd})/(f_0-f^\text{even})=\secondreview{1.45}\approx\sqrt{2}$, consistent with the expected spin-dependent tunneling amplitude.}

For device B, the situation changes significantly.
The energy ordering $E_0 > \ECS$ implies that quasiparticle states are not accessible (Fig.~1d).
We form a hybrid double dot by tuning T1, T2, and T3 close to pinchoff.
The CSD for a weak QD-SC coupling is shown in Fig.~3a.
The diagram is $2e$-periodic in the SC gate, indicating that the island is charged via Andreev reflections from the lead. 
The QD is again 1$e$-periodic.
To model the measured CSDs, we compute the charge ground state by diagonalizing an effective Hamiltonian of the system that includes charging effects, the superconducting gap in the island, and coupling terms \supp. 
This model, with the energy scales extracted from the Coulomb blockade measurements and an adjustable tunneling amplitude (rightmost panel in Fig.~3a), describes the observed CSD well.

The different gate charge periodicity for the QD and SC leads to interdot transitions that change the total charge of the dot-island system.
This implies that a reservoir must be involved in the corresponding charge-transfer process.
The observed resonator signal, with a linecut shown in Fig.~3b, results from tunneling on and off the QD, and thus should not contain information of SC-lead coupling \cite{esterli2018small}.
A possible candidate for the precise underlying process that gives rise to our data is crossed Andreev reflection (CAR) \cite{hofstetter2009cooper, herrmann2010carbon}.
There, a hole from the QD is converted to an electron in the lead, consistent with the charge states involved in the experiment.
This process is exponentially suppressed in the length of the island $\exp\left(-L/\pi\xi\right)$, where $\xi$ is the superconducting coherence length \cite{leijnse2013coupling}.
Still, with $L = \SI{1.2}{\micro\meter}$ and assuming a coherence length of $\xi \sim 260$\,nm \cite{albrecht2016exponential} this remains a plausible scenario.

Interestingly, increasing the tunnel coupling allows for bringing the system into a regime where a particle-conserving interdot transition emerges.
The CSD in a more strongly coupled regime, together with a simulation of the charge ground states is shown in Fig.~3c. 
In this regime, we assume an induced gap in the quantum dot, consistent with earlier studies on YSR states \cite{lee2014spin}.
Here, we observe that the regions with odd charge number in the QD shrink, while the regions with an even number of QD charges connect, resulting in an even-odd pattern in both gates.
Now, the interdot transition appears purely dispersive (Fig.~3d): we observe only a small phase shift, without any amplitude response.\sout{; this is indicative of a coherent transition.}
\new{Overall, our data is consistent with coherent Cooper pair transfer between the dot and the island.}
\sout{We can thus conclude that this transition is caused by coherent Cooper pair transfer between the dot and the island, resulting in an anti-crossing in the energy spectrum.}

\begin{figure}[t!]
	\includegraphics[width=0.5\textwidth]{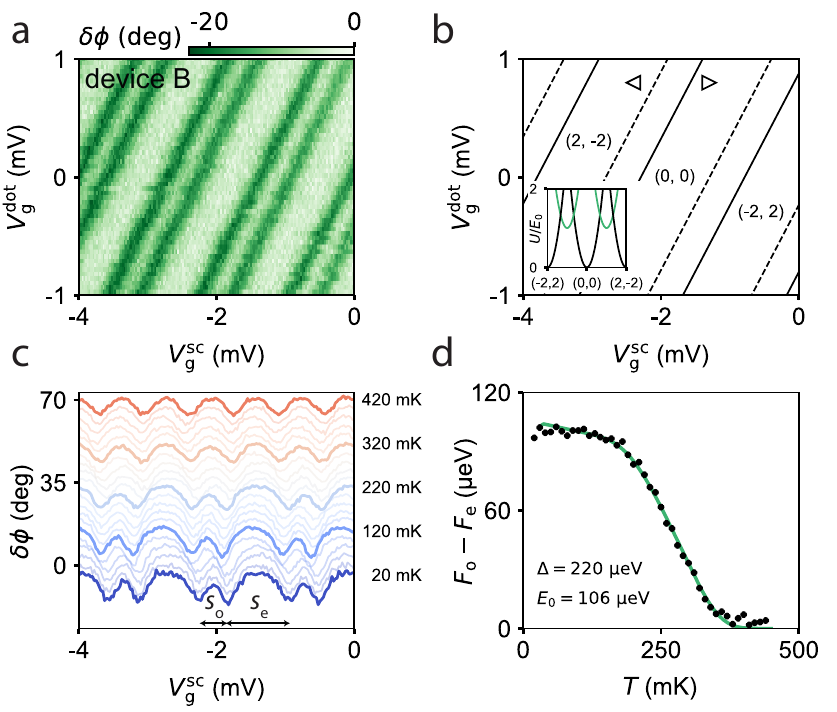}
	\caption{\textbf{\new{Spectroscopy of a subgap state in the floating regime}}
        \textbf{a)} 
        Charge stability diagram measured in phase in device B.
        The \sout{anti-}diagonal lines indicate that the total charge in the system is conserved.
        \textbf{b)} 
        Calculated positions of the transitions in good agreement with the measured stability diagram. 
        Inset: energy spectrum with even states in black and odd states in green, showing that the even-odd pattern is caused by the parity effect even though  $E_0>\ECS$.
        \textbf{c)} 
        Temperature dependence of the even-odd pattern.
        \textbf{d)} 
        The evolution of the free energy difference with temperature (black dots), with a fit to the model described in Ref. \citenum{higginbotham2015parity} (green line). 
        The free energy difference is extracted from the even-odd pattern via $F_\text{o}-F_\text{e} = \left(S_\text{e}-S_\text{o}\right)e\alpha_\text{sc}/4$ with $\alpha_\text{sc}=0.9$ the lever arm of the SC gate\sout{$G$, and $e$ the elementary electron charge}.
    }
	\label{Fig4}
\end{figure}

As we have seen, the main difference between the two devices is that the odd states of the SC can not be directly accessed in the regime $E_0 > \ECS$.
This changes in absence of lead reservoirs because quasiparticles that tunnel from the QD onto the SC are confined to the system \cite{esmail2017cooper}.
The additional energy associated with decharging the QD makes Cooper pair tunneling energetically unfavorable when $E_0<\ECS+\ECD$.
We realize this situation experimentally in device B by closing the outer tunnel barriers, through gates T1 and T3.
The resulting CSD and corresponding calculation of the ground state transitions are shown in Figs.~4a,b.
It can readily be seen that no transitions to a reservoir take place, and the even-odd pattern is indicative of the alternating occupation of even and odd states of the SC.

Importantly, even though SC and QD are now galvanically isolated from the environment, the gate sensor still allows us to study the quasiparticle states in the SC.
To show this, we measure the evolution of the even-odd spacing as a function of temperature (Fig.~4c).
This spacing is a measure for the free energy difference of the SC. 
In particular, the temperature evolution of the free energy difference can be used to identify and characterize subgap states \cite{lafarge1993measurement}; for proximitized nanowires, this has earlier been studied in transport \cite{higginbotham2015parity}.
The extracted free energy difference $F_\text{o}-F_\text{e}$ as a function of temperature is shown in Fig.~4d.
A fit to the model from Ref.~\citenum{higginbotham2015parity} yields a gap of $\Delta= \SI{220}{\micro\electronvolt}$, a supgap state energy of $E_0=\SI{106}{\micro\electronvolt}$, and an Al volume of $V=2.9\times10^5$ nm$^3$, consistent with the dimensions of the island.
We note that the slightly larger energy of the subgap state is consistent with the more negative plunger gate voltage for this measurement.
The excellent quality of the fit corroborates our initial assessment of the presence of a supgap state (Fig.~1b).
This result shows clearly that the resonator response of the QD gate sensor can be used to characterize states of the SC, even when leads for transport experiments are not available.

In summary, we have performed dispersive gate sensing on a quantum dot that can exchange particles with a superconducting island. 
Analysis of the resonator response has allowed us to directly detect the charge-tunneling processes that take place between the dot and the superconductor.
We have observed single or multi-particle tunneling processes, depending on the dominating energy scales of the hybrid double dot.
\new{Our results show that DGS provides an excellent tool for studying subgap excitations. 
In particular, using a QD and gate sensor allows performing spectroscopy without transport, which is relevant in cases where particle number should to be conserved, such as likely required for qubit devices that operate based on parity.
The ability of DGS to resolve differences in tunnel couplings --- as seen for the case of spin-dependent tunneling --- provides a very simple means for precisely characterizing hybridization while leaving the system in the ground state.}
Going forward, these demonstrated abilities will be crucial for the realization and operation of Majorana qubits based on proximitized nanowires \cite{karzig2017scalable, plugge2017majorana}.
Our results thus set the stage for the implementation of quantum measurements of topological qubits.

\section{\label{sec:acknowledgements}Acknowledgements}
We thank B. van Heck, K. Flensberg, A. Geresdi, A. Kou, and T. Karzig for useful discussions; and J. M. Hornibrook and D. J. Reilly for providing the frequency multiplexing chips. This work has been supported by the Netherlands Organization for Scientific Research (NWO), Microsoft, the Danish National Research Foundation, the European Research Council, and a Justus and Louise van Effen excellence scholarship.

\end{document}